\documentclass[usenatbib,useAMS]{mn2e}

\usepackage[dvips]{graphicx}
\usepackage{dcolumn}
\usepackage{times}

\title{B0850+054: a new gravitational lens system from CLASS}
\author[A.~D.~Biggs et al.]{A.~D.~Biggs,$^{1,2}$\thanks{E-mail:
    biggs@jive.nl} D.~Rusin,$^{3,4}$ I.~W.~A. Browne,$^1$ A.~G.~de
    Bruyn,$^{5,6}$ N.~J.~Jackson,$^1$ 
\newauthor L.~V.~E.~Koopmans,$^7$ J.~P.~McKean,$^1$ S.~T.~Myers,$^8$
R.~D.~Blandford,$^7$ K.~-H.~Chae,$^1$
\newauthor C.~D.~Fassnacht,$^9$ M.~A.~Norbury, $^1$ T.~J.~Pearson,$^7$
P.~M.~Phillips,$^1$ A.~C.~S.~Readhead$^7$
\newauthor and P.~N.~Wilkinson$^1$\\
$^1$University of Manchester, Jodrell Bank Observatory, Macclesfield, Cheshire 
SK11 9DL\\
$^2$Joint Institute for VLBI in Europe, Postbus 2, 7990 AA Dwingeloo,
The Netherlands\\
$^3$Department of Physics and Astronomy, University of Pennsylvania,
209 S. 33rd St., Philadelphia, PA 19104-6396, USA\\
$^4$Harvard-Smithsonian Center for Astrophysics, 60 Garden Street,
Cambridge, MA 02138, USA\\
$^5$Kapteyn Astronomical Institute, Postbus 800, 9700 AV Groningen, The Netherlands\\
$^6$ASTRON, Postbus 2, 7990 AA Dwingeloo, The Netherlands\\
$^7$California Institute of Technology, Pasadena, CA 91125, USA\\
$^8$National Radio Astronomy Observatory, P.O. Box 0, Socorro, NM
87801, USA\\
$^9$Space Telescope Scientific Institute, 3700 San Martin Dr.,
Baltimore, MD 21218, USA}
\begin{document}
\maketitle
\begin{abstract}
We report the discovery of a new gravitational lens system from the
CLASS survey. Radio observations with the VLA, the WSRT and MERLIN show
that the radio source B0850+054 is comprised of two compact
components with identical spectra, a separation of 0.7~arcsec and a
flux density ratio of 6:1. VLBA observations at 5~GHz reveal structures
that are consistent with the gravitational lens hypothesis. 
The brighter of the two images is resolved into a linear string of at
least six sub-components whilst the weaker image is radially
stretched towards the lens galaxy. UKIRT $K$-band imaging detects an
18.7~mag extended object, but the resolution of the observations is not
sufficient to resolve the lensed images and the lens galaxy. Mass
modelling has not been possible with the present data and the
acquisition of high-resolution optical data is a priority for this
system. 
\end{abstract}


\section{Introduction}

The Cosmic Lens All-Sky Survey (CLASS) \citep{myers02,browne02} is a
survey of flat-spectrum sources with a flux density in excess of 30~mJy
at 5~GHz. 22 lenses have been found in CLASS from a total of 16,503
sources observed. The main aim of this survey is to identify
gravitational lens systems. These can be used to constrain the
cosmological parameters $H_0$
\citep[e.g.][]{biggs99,fassnacht99,koopmans00}, $\Omega_0$ and
$\lambda_0$ \citep[e.g.][]{kochanek96,helbig99,chae02} as well as
galaxy mass profiles \citep[e.g.][]{rusin01,cohn01}.  

CLASS B0850+054 has a 5-GHz GB6 \citep{gregory96} flux density of
68~mJy and a two-point spectral index\footnote{We define the spectral
  index, $\alpha$, such that $S_{\nu} \propto \nu^{\alpha}$.} between
1.4 and 5~GHz of $\alpha=-0.2$, thus satisfying the criteria for
inclusion in the CLASS sample. The original 8.5-GHz VLA snapshot (30-s
on source) revealed two compact components separated by 0.7~arcsec,
with a flux density ratio of $\sim$6:1. Here we present follow-up
observations of B0850+054 in the radio, optical and infra-red which
leave little doubt that B0850+054 is a lens system.

\section{Radio observations}
\label{radiosec}

B0850+054 has been observed with the VLA,
the Multi-Element Radio-Linked Interferometer Network (MERLIN), the
Very Long Baseline Array (VLBA) and the Westerbork Synthesis Radio
Telescope (WSRT). A summary of the observations presented in this
paper is given in Table~\ref{obstab}.

\subsection{VLA}
\label{vlasec}

The VLA data consist of observations at 5, 8.5 and 15~GHz. At each
frequency two channels of 50-MHz bandwidth were used and the target
observed for about 10~min. Initial calibration was done with the NRAO
{\sc aips} package and the data were  
then mapped and self-calibrated in {\sc difmap} \citep{shepherd97}. The
resultant maps are shown in Fig.~\ref{vlamerfig}. These show that the
source consists of two compact components aligned approximately
north--south. As is customary, the brighter of the two components is
denoted image A and the other image B. The results of model-fitting to
the ($u,v$) data in {\sc difmap}, using an unresolved point (delta)
component to represent each image, are presented in
Table~\ref{modtab}. We also tabulate the flux density ratios,
demonstrating that the spectral index of each component is the same.

\begin{table}
\begin{center}
\caption{Radio observations of B0850+054 presented in this paper.}
\begin{tabular}{lll} \\ \hline
Date & Array & $\nu$ (GHz)   \\ \hline
1998 Mar 15 & VLA    & 8.5   \\
2000 Oct 07 & VLBA   & 5.0   \\
2001 Jan 27 & MERLIN & 5.0   \\
2002 Feb 04 & WSRT   & 0.360 \\
2002 Apr 14 & VLA    & 5.0   \\
2002 Apr 14 & VLA    & 8.5   \\
2002 Apr 14 & VLA    & 15.0  \\ \hline
\end{tabular}
\label{obstab}
\end{center}
\end{table}

\subsection{MERLIN}
\label{merlinsec}
\begin{figure*}
\begin{center}
\includegraphics[scale=0.42]{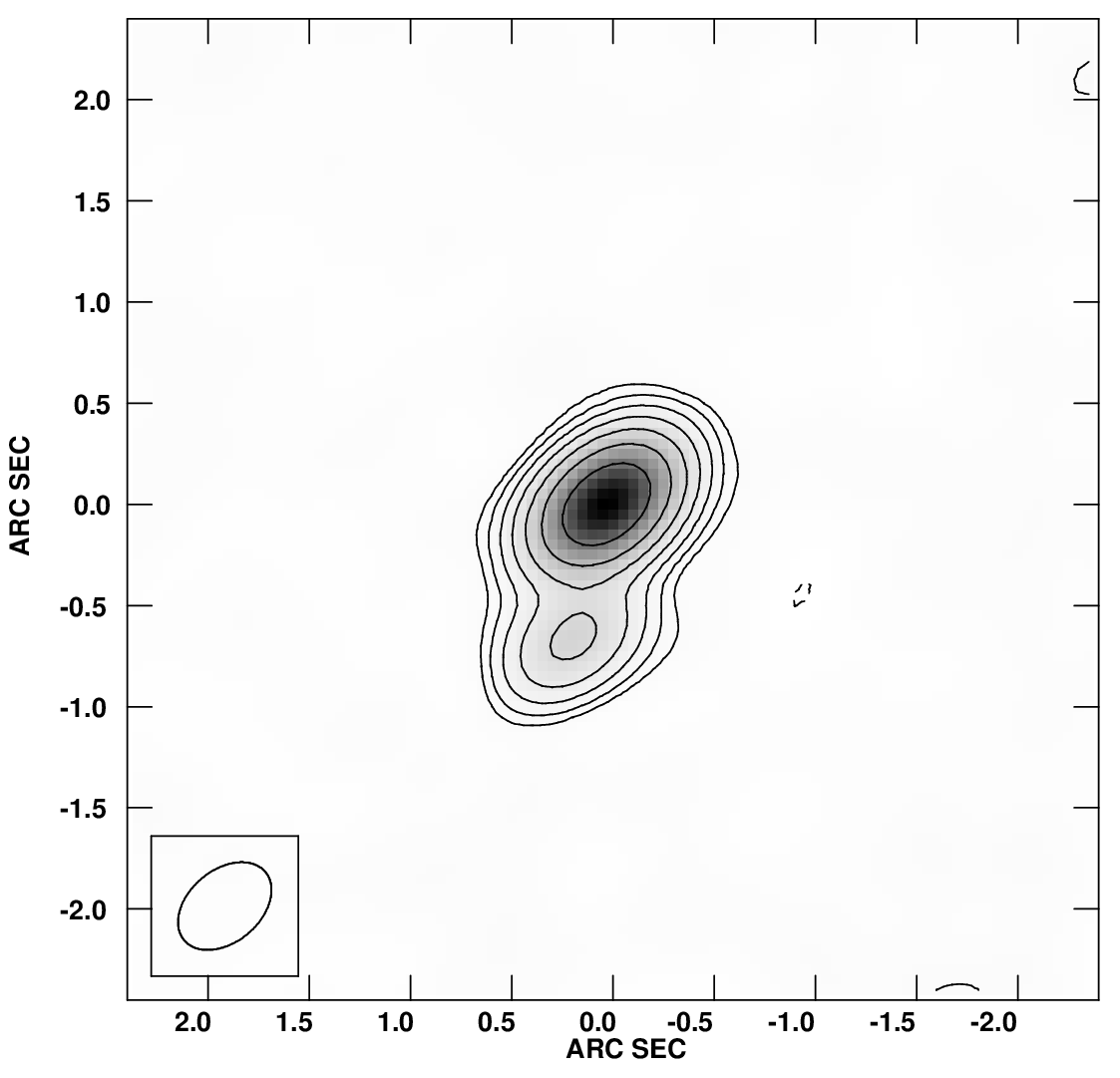}
\includegraphics[scale=0.42]{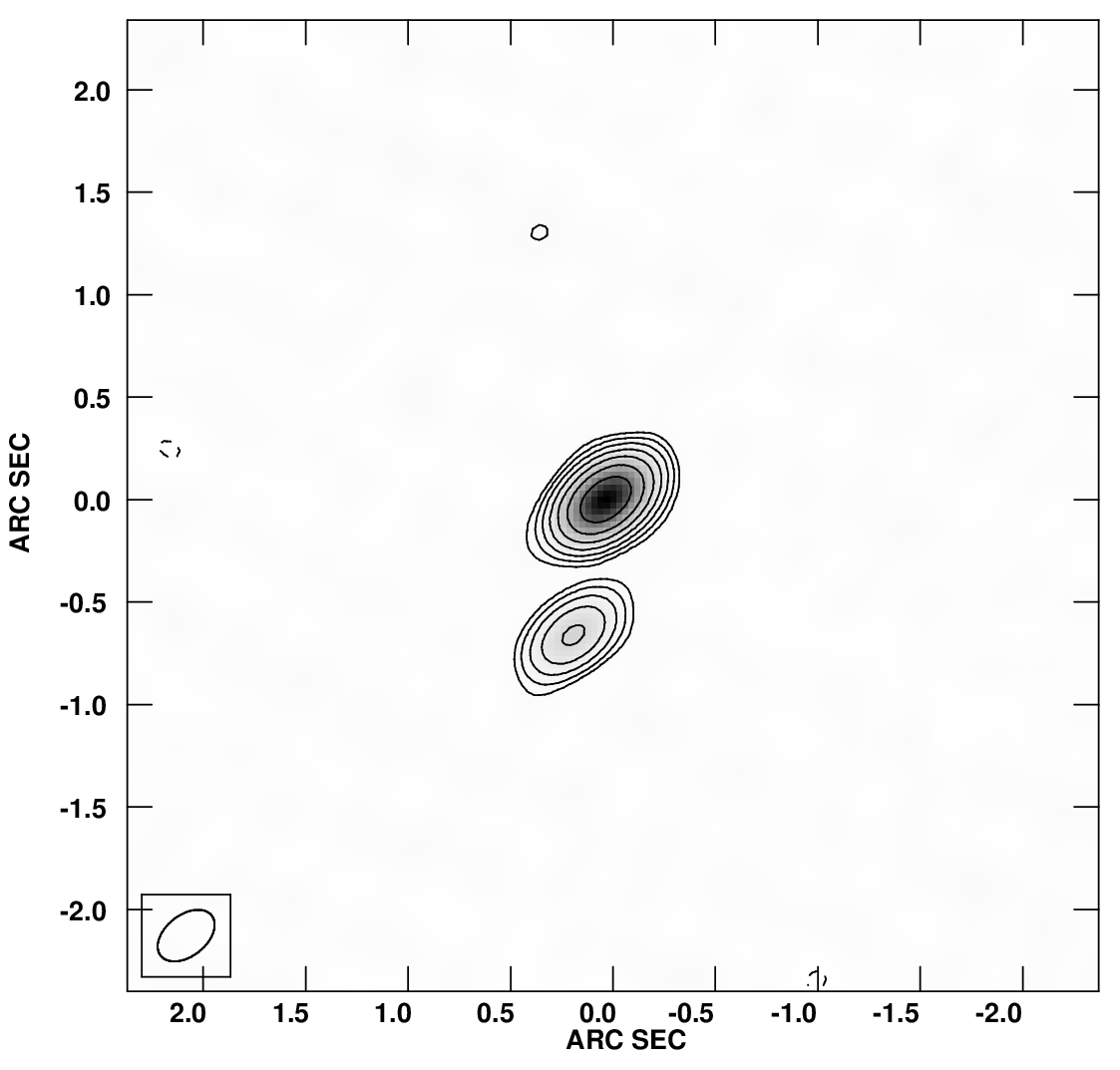}
\includegraphics[scale=0.42]{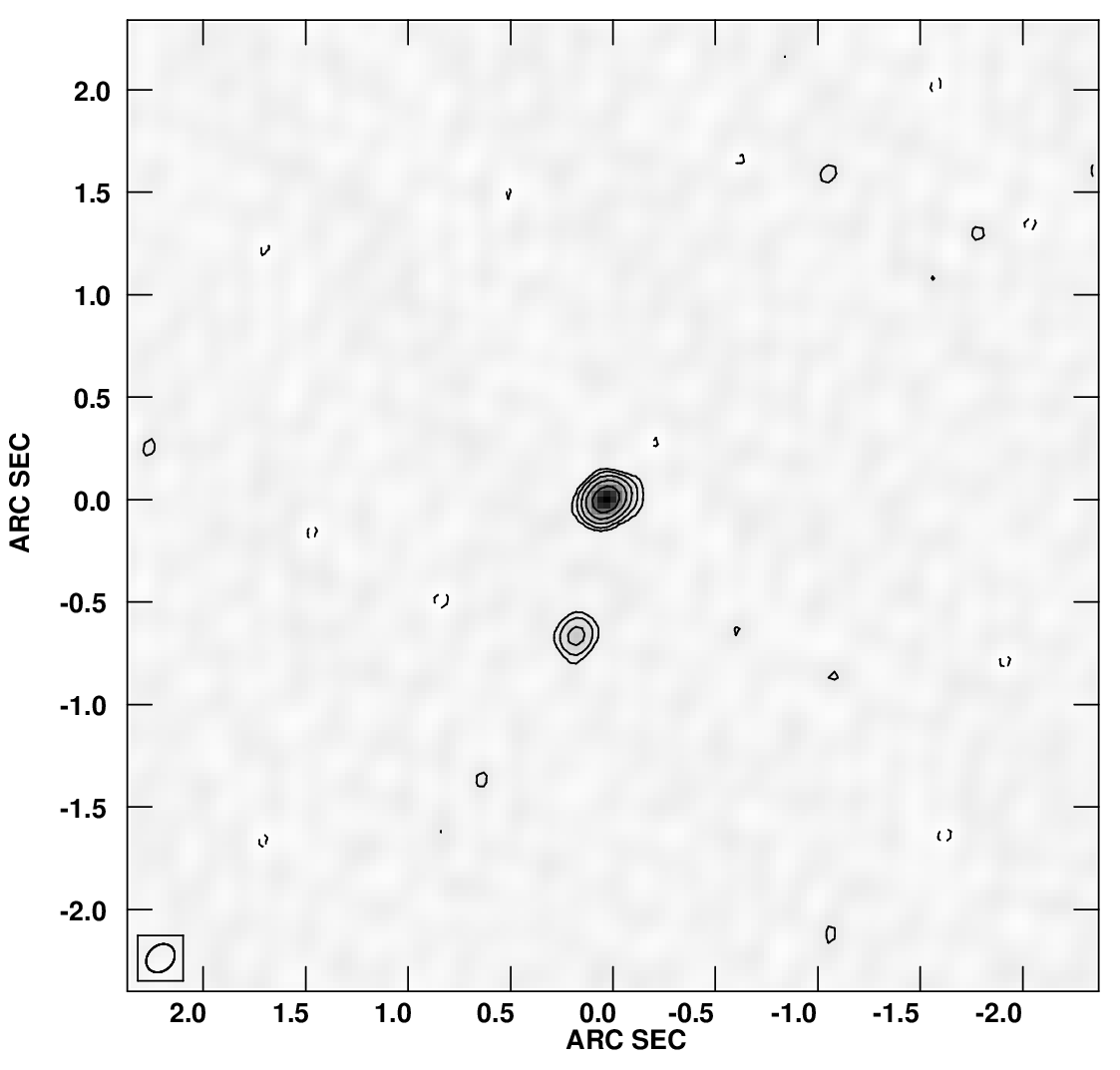}
\includegraphics[scale=0.42]{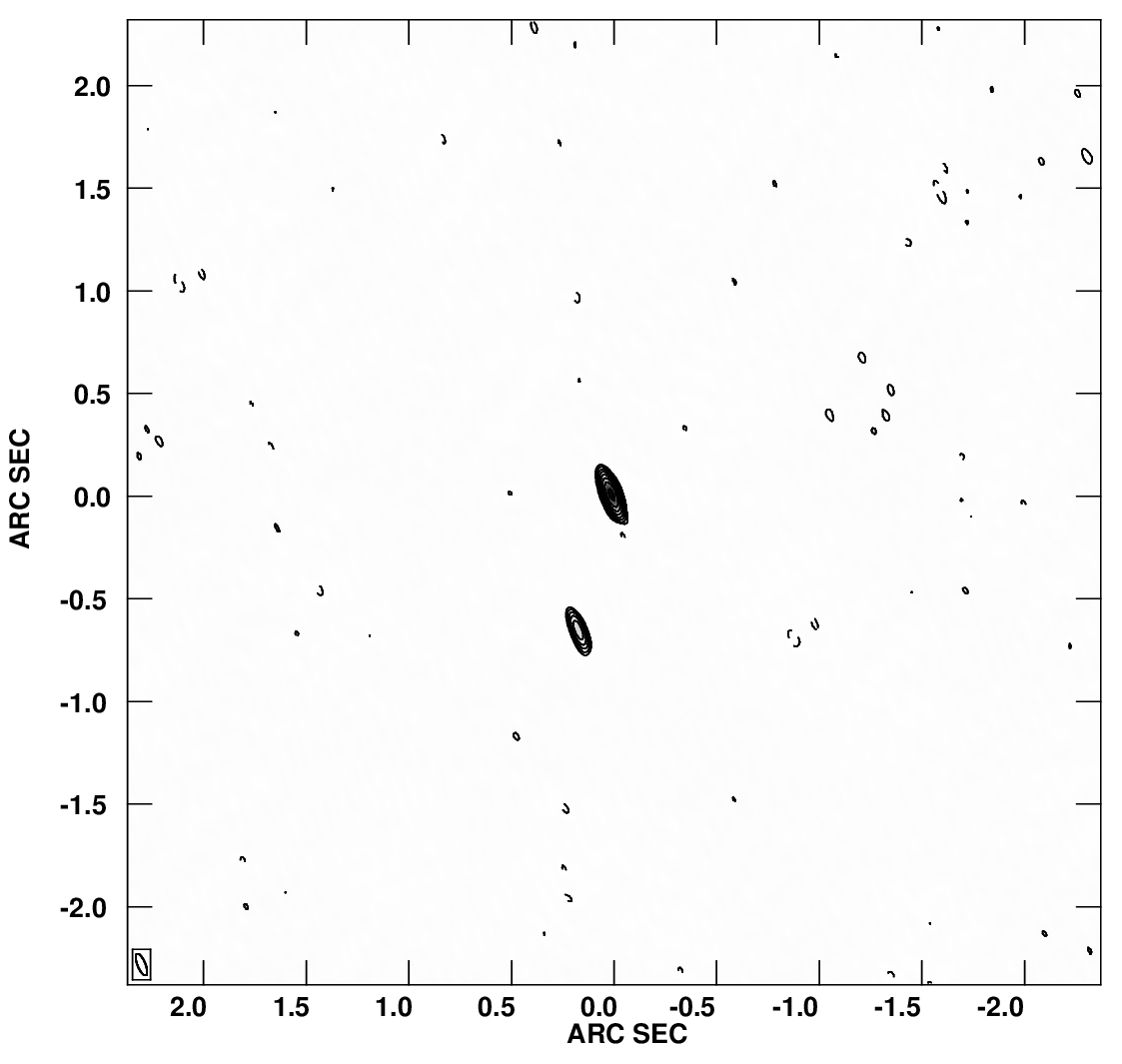}
\caption{VLA and MERLIN maps of B0850+054. Top left: VLA 5-GHz map with
  a restoring beam of $524 \times 356$~mas at a position angle 
  of $-49$\fdg9. The data are uniformly weighted and the rms noise
  in the map is 157~$\mu$Jy~beam$^{-1}$. Top right: VLA 8.5-GHz map with
  a restoring beam of $317 \times 200$~mas at a position angle 
  of $-51$\fdg6. The data are uniformly weighted and the rms noise
  in the map is 121~$\mu$Jy~beam$^{-1}$. Bottom left: VLA 15-GHz map with
  a restoring beam of $153 \times 123$~mas at a position angle 
  of $-46$\fdg2. The data are uniformly weighted and the rms noise
  in the map is 288~$\mu$Jy~beam$^{-1}$. Bottom right: MERLIN 5-GHz map
  with a restoring beam of $114 \times 38$~mas at a position angle 
  of $22$\fdg2. The data are naturally weighted and the rms noise in
  the map is 100~$\mu$Jy~beam$^{-1}$. All maps are plotted on the
  same angular scale and each has the restoring beam plotted in the
  bottom-left corner. Contours are plotted at multiples ($-1$, 1, 2, 4,
  8, 16, etc) of 3$\sigma$ where $\sigma$ is the off-source rms noise
  in the map. Image A is the northern of the two components.}
\label{vlamerfig}
\end{center}
\end{figure*}

The MERLIN 5-GHz observations were taken over a period of 9~h, with a
bandwidth of 16~MHz, and were phase-referenced. The data were 
calibrated in {\sc aips} and mapped and self-calibrated in {\sc
  difmap}; a map made from naturally-weighted data is shown in
Fig.~\ref{vlamerfig}. As with the VLA data we modelfit using {\sc
  difmap} and show the results in Table~\ref{modtab}. The higher
resolution of MERLIN allows a more accurate estimation of the image
separation ($679\pm1$~mas) and also allows us to detect signs of
resolution in image A. For this reason we fit an elliptical Gaussian to
image A whilst still modelling image B with a delta component. No
evidence for polarisation of the images has been found down to the
noise limit. This implies that the polarisation of the lensed radio
source is $<1$~per~cent ($5~\sigma$) at 5~GHz.

\begin{table*}
\begin{center}
\caption{Results of model-fitting to the MERLIN and VLA data. The VLA data
  were model-fitted assuming delta components for both A and B whilst
  the MERLIN data were fitted with an elliptical Gaussian to A and a
  delta component to B. Included are the flux densities of images A and
  B, their flux ratio, their separation ($r$) and their orientation
  ($\theta$) measured in degrees east from north. We adopt errors of
  3~per~cent on the VLA 5 and 8.5-GHz flux densities and 5~per~cent on
  the MERLIN 5-GHz and VLA 15-GHz.}
\begin{tabular}{lllllll} \\ \hline
$\nu$ (GHz) & Array & $S_{\nu,{\mathrm{A}}}$ (mJy) & $S_{\nu,{\mathrm{B}}}$
(mJy) & $S_{\nu,{\mathrm{A}}}/S_{\nu,{\mathrm{B}}}$ & $r$ (mas) & $\theta$ ($^{\circ}$) \\ \hline
5   & MERLIN & 49.0 & 8.1 & $6.0\pm0.4$ & $679\pm1$ & $166.5\pm0.1$ \\
5   & VLA    & 55.1 & 9.0 & $6.1\pm0.3$ & $683\pm8$ & $166.2\pm0.7$ \\
8.5 & VLA    & 40.4 & 6.6 & $6.1\pm0.3$ & $679\pm5$ & $166.6\pm0.4$ \\
15  & VLA    & 26.2 & 4.4 & $6.0\pm0.4$ & $680\pm8$ & $166.6\pm0.7$ \\ \hline
\end{tabular}
\label{modtab}
\end{center}
\end{table*}

\subsection{VLBA}
\label{vlbasec}

\begin{figure*}
\begin{center}
\includegraphics[scale=0.42]{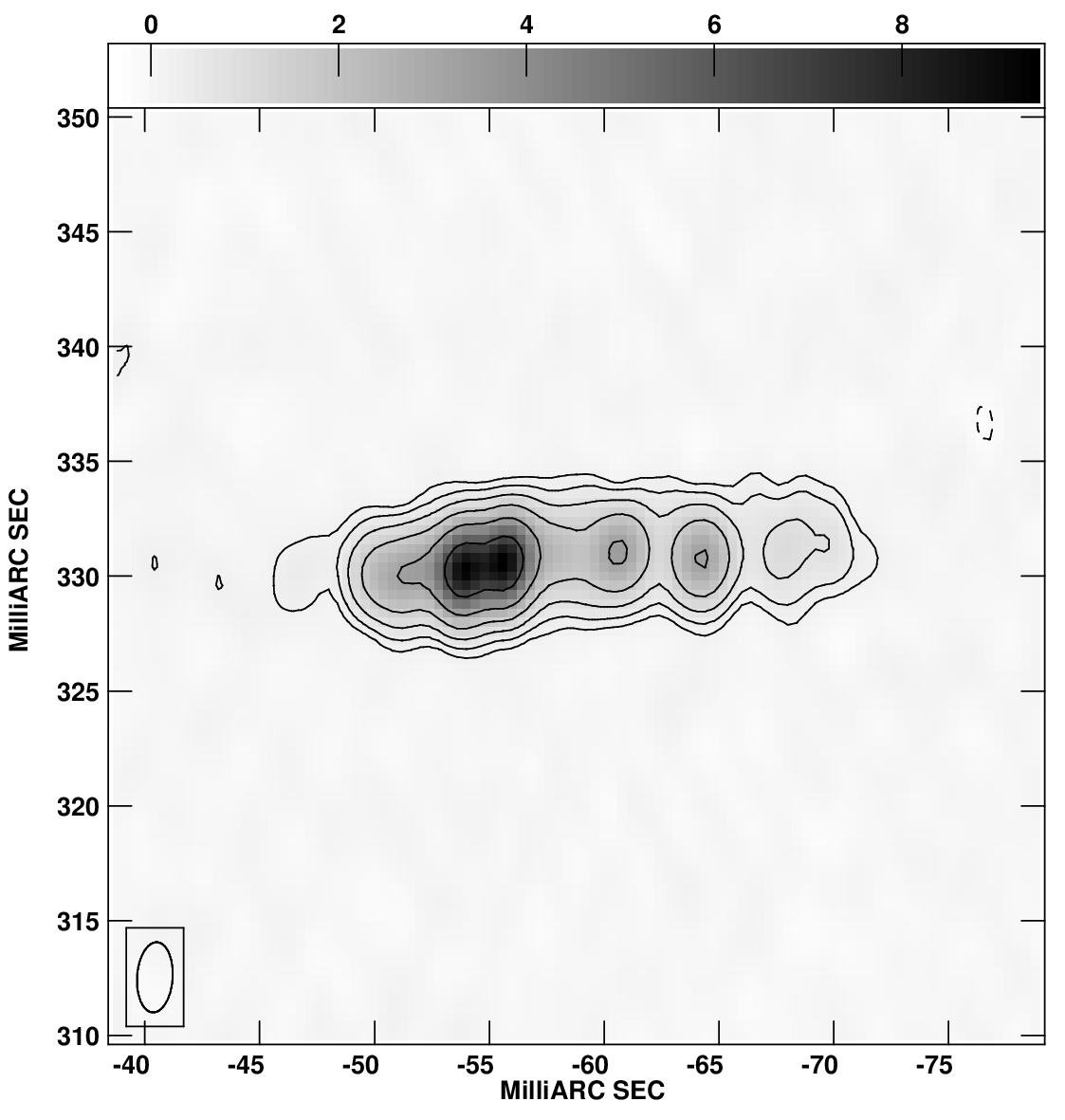}
\includegraphics[scale=0.42]{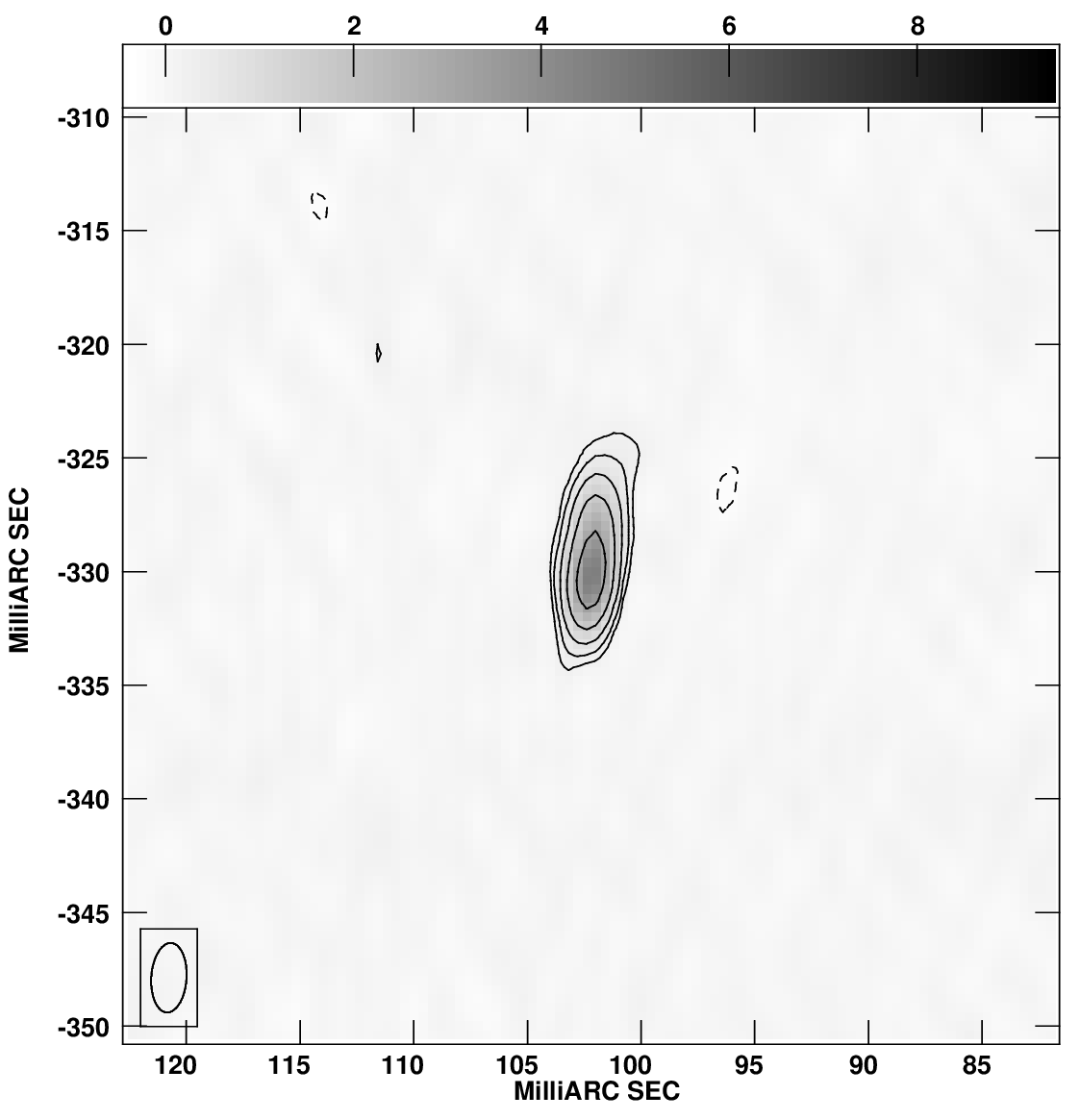}
\caption{VLBA 5-GHz maps of image A (left) and image B (right) of
  B0850+054. The maps are uniformly weighted and both are plotted on
  the same angular scale. The 
  restoring beam is shown in the bottom-left of each map and has a FWHM
  of $3.1 \times 1.6$~mas at a position angle of $-3$\fdg8. Contours
  are plotted at multiples ($-1$, 1, 2, 4, 8, 16, etc) of 3$\sigma$
  where $\sigma$ is the off-source rms noise in the
  map (50~$\mu$Jy~beam$^{-1}$). The grey scale represents surface
  brightness in units of mJy~beam$^{-1}$. Positions are offset from
  $08^{\rmn{h}} 52^{\rmn{m}} 53\fs5772, +5\degr 15\arcmin 15\farcs329$
  (J2000).}
\label{vlbafig}
\end{center}
\end{figure*}

Phase-referenced observations were taken with the VLBA at a frequency
of 5~GHz for a total of 12~h. A total bandwidth of 32~MHz was observed
in four separate, but contiguous, bands of 8~MHz each. The data rate
was 128~Mb~s$^{-1}$. Calibration and mapping took place within {\sc
  aips} following standard procedures. Naturally-weighted maps of
images A and B of 0850+054 are shown in Fig.~\ref{vlbafig}. 
Image A is found to have a total length of about
20~mas and consists of a bright ``core'' region (containing two
sub-components of similar brightness) near to its eastern extremity and
several bright knots to the west of this. To the east of these
sub-components there is a short extension which is most likely a
counterjet. As expected for a de-magnified version of A, image B is
relatively compact. It is, however, resolved north-south suggesting
that with higher resolution the sub-components in the jet could be
mapped.

\subsection{WSRT}
\label{wsrtsec}

In order to obtain a low-frequency flux density measurement of
B0850+054, observations were made with the WSRT with its broadband
``92-cm'' system for two periods of 2~h, separated by about 4~h.
Three of the eight 5-MHz bands were affected by RFI and
discarded. After mapping and self-calibrating we detected a 
source of $50\pm5$~mJy close to the position of the target. Due to the
very elongated beam at this declination the possibility of confusion
with unrelated sources is significant. After searching the 1.4-GHz
NVSS \citep{condon98} and FIRST \citep*{becker95} surveys, both of
which detect B0850+054, we identified a $\sim$14~mJy confusing source
22~arcsec west and 144~arcsec north of B0850+054, well within the WSRT
beam. As this source is also detected in our VLA data at 5 and 8.5~GHz,
together with the 1.4-GHz flux densities we are able to calculate its
spectral index, $\alpha = -0.8$.

Assuming that the spectrum of this source does not turn over, its flux
density at 360~MHz would be $\sim$41~mJy, thus suggesting that it is
contributing significantly to the measured WSRT flux density. We have
aligned the Right Ascension reference frame of the WSRT 360-MHz image
with that of the NVSS using six sources that are common to both. The
scatter in this relative astrometry is less than 1~arcsec and from it
we have determined that the WSRT detection lies approximately
midway between the NVSS positions of the two sources. Therefore we
conclude that each contribute about equally to the combined flux
density and adopt a value of $25\pm5$~mJy for B0850+054 at 360~MHz.

\section{UKIRT observations} 

\begin{figure}
\begin{center}
\includegraphics[scale=0.38,angle=0]{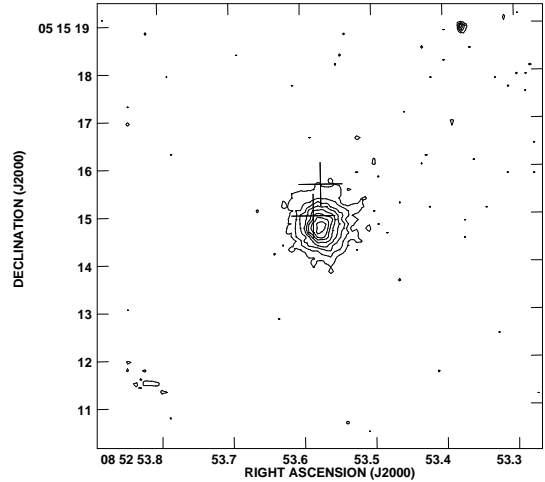}
\caption{UKIRT $K$-band image of B0850+054. Contours are plotted at the
  background sky level plus linearly-increasing multiples ($-1$, 1, 2,
  3, etc) of three times the off-source rms noise. The crosses mark the 
  position of the A and B components.}
\label{ukirtfig}
\end{center}
\end{figure}

Service observations with the United Kingdom Infra-Red Telescope
(UKIRT) in $K$ band (2.2~$\mu$m) were taken on the nights 
of 2001 March 17 and 2001 March 18 using the UKIRT Fast-Track Imager
(UFTI). Both nights' observations were made using a mosaic mode, each
mosaic consisting of nine 60~s pointings with an offset between each of
10~arcsec. A total of nine mosaics were obtained over the two
nights. Short observations (20~s) of the standard star FS15 were taken
on each night so that the data could be flux calibrated. Data
reduction was performed using the UKIRT pipeline analysis tool,
ORAC-DR. The $K$-band image resulting from the sum of all mosaics is
shown in Fig.~\ref{ukirtfig}. 

We detect an 18.7~mag source at the location of the radio images, but
unfortunately the seeing is not adequate to resolve the radio images.
From measurements of the sizes of stars in the field we measure a
seeing-disc size of about 0.6~arcsec, only marginally less than the
separation between the radio components. The source is however extended,
with a major axis (FWHM~$\sim~$0.8--0.9~arcsec) that is aligned
approximately north--south ($\theta\sim10\degr$). As this is similar
to the orientation of the lensed images it is likely that some lensed
emission has been detected. We have attempted to accurately align the
radio and UKIRT images by using the position of one of the stars in the
UKIRT field that has been measured with the Carlsberg Automatic
Meridian Circle (D.~W.~Evans, private communication). The positions of
the radio components are also plotted in Fig.~\ref{ukirtfig}. The
accuracy of our astrometry is set by the positional accuracy of the
Carlsberg star which, with a positional uncertainty of about 150~mas,
means that we cannot be certain if most of the emission we see is from
the lens or the lensed images.

\section{Discussion}

We have presented radio and infra-red observations of the CLASS source
B0850+054. In this section we describe the evidence that leads us to
believe that this source is in fact a doubly-imaged lensed system.

\begin{figure}
\begin{center}
\includegraphics[scale=0.34,angle=-90]{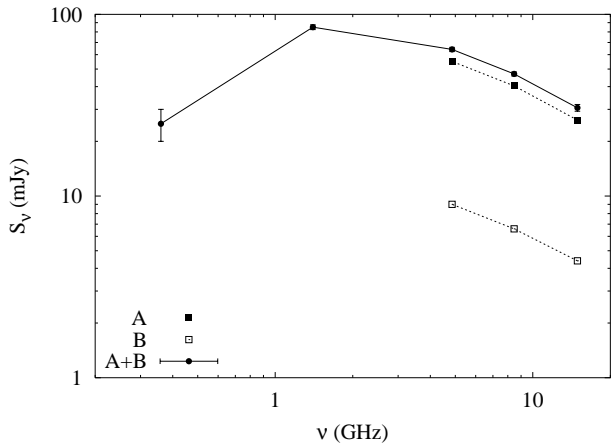}
\caption{Radio spectrum of B0850+054. The solid line shows total
  flux densities (A+B) whilst below this the individual A and B flux
  densities are plotted where possible.}
\label{radiospec}
\end{center}
\end{figure}

The most convincing evidence is provided by the radio data. Our VLA,
MERLIN and WSRT measured flux densities at 15, 8.5, 5 and 0.325~GHz,
plus the 1.4-GHz NVSS measurement are plotted as a function of
frequency in Fig.~\ref{radiospec}. Below 5~GHz the existing maps do 
not have the resolution to enable us to separate the A and B images and
so only a  total A+B flux density is plotted. At frequencies $\ge$5~GHz
the excellent agreement between the radio spectra of A and B is clear
and the flux ratios in Table~\ref{modtab} confirm that the two
components have identical spectra to within the errors.

Further support for the lensing hypothesis comes from a consideration
of surface brightness arguments. As a lens system preserves surface
brightness, the weaker image must be the more compact as demonstrated
particularly well by the VLBA data. Moreover, the extension of B along
the line of the A-B separation is characteristic of gravitational
lensing and the lack of detectable polarisation in each image is also
consistent with the lensing hypothesis. 

A spectrum taken with the W.~M.~Keck~I telescope has measured a
redshift for the lensing galaxy of $z_l =0.59$ (McKean et al., in
preparation), based on the detection of Ca H \& K and G-band absorption
as well as a 4000~{\AA} break. Also detected is a single emission
line which is almost certainly associated with the lensed radio
source. The most likely identification for this line is
\mbox{Mg\,{\sc ii}} ($z_s=1.14$) though it could possibly be Ly$\alpha$
($z_s=3.93$). Assuming that \mbox{Mg\,{\sc ii}} is the correct
identification, the image separation together with the measured lens
redshift and a source redshift of 1.14 predicts a $K$ magnitude for the
lens galaxy of 17.9,\footnote{Invoking the Faber-Jackson relation
  \citep{faber76} and assuming that an $L^{\ast}$ galaxy at $z=0.6$ has
  a $K$ magnitude of 16.4.} close to the measured UKIRT magnitude of
18.7. Unfortunately, the present optical data do not provide enough
constraints (in particular, the galaxy position relative to the lensed
images is unknown) to make any attempt at mass modelling fruitful.

From Fig.~\ref{radiospec} it can be seen that the source spectrum has a
turn-over at approximately 1~GHz, thus making it a member of the
giga-hertz peaked spectrum (GPS) class. Low flux density variability is
a generally recognised characteristic of GPS sources, as is low
polarisation \citep{odea98} and so B0850+054 seems consistent with the
GPS identification. The source is $<1$~per~cent polarised, and would
also appear to be only moderately variable, if at all. Flux density
measurements from the VLA and MERLIN data, the GB6 survey and the
Parkes-MIT-NRAO (PMN) survey \citep{griffith95}, all at 5~GHz, are
identical to within their errors. At 8.5~GHz, recent monitoring with
the VLA (six epochs over three months) found no evidence for
significant variability. Furthermore, an intrinsic size for the GPS
source of $\sim$10~mas (as measured from the VLBA map of image A and
corrected for a lens magnification of $\sim$2, a plausible value) is
consistent with the frequency and flux density (also corrected for the
lens magnification) of the turnover using the relation found by
\citet{snellen00} for GPS sources.

\section{Summary and future work}

We have discovered a new lens system during the course of the CLASS
survey, B0850+054. It consists of two images of a GPS radio source with a
separation of 0.7~arcsec and a flux density ratio of $\sim$6:1. UKIRT
$K$-band imaging does not resolve the lens galaxy and the lensed
images, but a lens galaxy redshift of 0.58 has been measured with the
W.~M.~Keck telescope.

A priority for future work is the acquisition of a high-resolution
optical/infra-red image either with the $HST$ or with ground-based
adaptive optics. Knowing the relative positions of the
lens and the lensed images will enable detailed mass modelling to be
undertaken. We also plan higher resolution VLBI observations to look
for the substructure in image B that must be there. At the same time
we hope to detect motion of the sub-components. There is the exciting
possibility, if B0850+054 is really a two-sided object, of being able
to measure the advance speeds of the approaching and receding
jets, made more easily visible by the lens magnification. When combined
with measurement of the relative flux densities of the receding and
approaching components, this opens up the prospect of an independent
constraint on the Hubble constant \citep[e.g.][]{taylor97}.

\section*{Acknowledgments}

JPM, MAN and PMP acknowledge the receipt of PPARC studentships. LVEK
acknowledges grants from the National Science Foundation and NASA (AST
99-00866, STScI-GO 06543.03-95A and STScI-AR-09222). The VLA
and the VLBA are operated by the National Radio Astronomy Observatory
which is a facility of the NSF operated under
cooperative agreement by Associated Universities, Inc. MERLIN is run by
the University of Manchester as a National Facility on behalf of
PPARC. The WSRT is operated by ASTRON (Netherlands Foundation for
Research in Astronomy) with support from the Netherlands Foundation for
Scientific Research (NWO). UKIRT is operated by the Joint Astronomy 
Centre on behalf of PPARC. We would like to thank the support staff of
UKIRT with their help with the observations and Dafydd Wyn Evans for
his help with the UKIRT astrometry. The W.~M.~Keck Observatory is
operated as a scientific partnership among the California Institute of
Technology, the University of California and NASA. This research was
supported in part by the European Commission TMR Programme, Research
Network Contract ERBFMRXCT96-0034 `CERES'. We appreciate the comments
of the anonymous referee.

\bibliographystyle{mnras}
\bibliography{0850}

\end{document}